\documentclass{aastex}          
\usepackage{spr-astr-addons}    

\begin{document}
%
\title{An HST/NICMOS view of the prototypical giant H\,II region NGC\,604 in
  M33} 

\shorttitle{NICMOS view of NGC\,604}
\shortauthors{Barb\'a et al.}

\author{Rodolfo H. Barb\'a\altaffilmark{1}} 
\affil{Departamento de F\'{\i}sica, Universidad de La Serena, Benavente 980,
  La Serena, Chile}
\email{rbarba@dfuls.cl}
\author{Jes\'us Ma\'{\i}z Apell\'aniz\altaffilmark{2}}
\author{Enrique P\'erez}
\affil{Instituto de Astrof\'{\i}sica de Andaluc\'{\i}a-CSIC, Camino bajo de
  Hu\'etor 50, Granada, Spain}
\author{M\'onica Rubio\altaffilmark{}}
\affil{Departamento de Astronom\'{\i}a, Universidad de Chile, Santiago, Chile}
\author{Alberto Bolatto\altaffilmark{3}}
\affil{Dept. of Astronomy and Laboratory for Millimeter-Wave Astronomy,
  University of Maryland, College Park, MD 20742, USA}
\author{Cecilia Fari\~na\altaffilmark{4}}
\author{Guillermo Bosch\altaffilmark{1,4}}
\affil{Facultad de Ciencias Astron\'omicas y Geof\'{\i}sicas, Universidad
  Nacional de La Plata, Argentina}
\author{Nolan R. Walborn\altaffilmark{}}
\affil{Space Telescope Science Institute, 3700 San Martin Dr., Baltimore, MD 21218, USA}

\altaffiltext{1}{Member of the Carrera del Investigador Cient\'{\i}fico
  CONICET, Argentina} 
\altaffiltext{2}{Ram\'on y Cajal Fellow}
\altaffiltext{3}{Department of Astronomy and Radio Astronomy Laboratory,
  University of California, Berkeley, CA 94720, USA}
\altaffiltext{4}{Instituto de Astrof\'{\i}sica La Plata, CONICET, La Plata, Argentina}

\begin{abstract}
We present the first high-spatial resolution near-infrared (NIR) imaging of
\object{NGC\,604}, obtained with the NICMOS camera aboard the Hubble Space Telescope (HST). 
These NICMOS broadband images reveal new NIR point sources, clusters, and 
diffuse structures. We found an excellent spatial correlation between the
8.4 GHz radio continuum and the 2.2\,$\mu$m nebular emission. Moreover, 
massive young stellar object candidates appear aligned with these radio peaks, 
reinforcing the idea that those areas are star-forming regions. 
Three different scaled OB associations are recognized in the 
NICMOS images. The brightest NIR sources in our images have properties
that suggest that they are red supergiant stars, of which one of them was
previously known. 
This preliminary analysis of the NICMOS images shows the complexity of the
stellar content of the NGC\,604 nebula. 
\end{abstract}

\keywords{galaxies: starburst -- HII Regions -- ISM: individual (NGC\,604) --
  Messier 33 -- stars: red supergiants -- stars: early-type -- stars: formation}

%
\section{Motivation for a detailed study of NGC\,604}\label{s:intro}
Giant \ion{H}{2} regions (GHRs) are among the most luminous objects that can
be individually identified very distant galaxies. In these regions star
formation occurs at extremely high rates, hence they are also known as
starburst regions. Given the young age and large number of massive stars in
GHRs, some of them are still found embedded in their dusty parental molecular
clouds. The morphology of these regions changes quite rapidly during the
first few million years after the first massive generation is
born. Excellent pictures showing these evolutionary morphologies are the
\ion{H}{2} regions in the irregular galaxy \object{NGC\,4214}
\citep{maiz2000,mack2000}.

The natural extension to larger scales are the starburst galaxies which can be
observed at cosmological distances and, therefore, can be used as astrophysical
signposts to trace the star formation history of the Universe. However, due to
their large distances, we must rely only on their major global properties, 
such as the behavior of the strong recombination lines originating in the 
ionized nebula.
In order to improve our knowledge of these objects in distant galaxies, 
it is necessary to study resolved nearby examples. 
\object{30 Doradus} in the \object{LMC} and \object{NGC\,604} in \object{M33}
are the two largest GHRs in the Local Group (LG), so they are the prime
candidates for such study.  

\object{NGC\,604} is 
located at a distance of 840\,kpc \citep{freedman2001}. 
At such a distance, $1\arcsec$ is equivalent to 4\,pc and consequently  
the capabilities of HST are essential to
study the region with high spatial resolution. 
\object{NGC\,604} is powered by a massive young cluster without a central core 
that contains over 200 O and WR stars (\citeauthor{maiz2001} 
\citeyear{maiz2001} and references therein). This GHR is  
the best nearby example of a scaled OB association, or SOBA, as defined by 
\cite{maiz2001}, 
a more extended type of object than \object{30\,Dor},
which is a Super Star Cluster (SSC). The combination of similarities and
differences between those objects makes the combined analysis of both 30 Doradus and
NGC\,604 a
necessary step for the creation of a template for the understanding of
far-away starbursts.

In this contribution we describe the gathered observations and general
objectives, and we present a preliminary analysis of the new HST/NICMOS
images. 

%
\begin{figure}[t]
\includegraphics[width=\columnwidth]{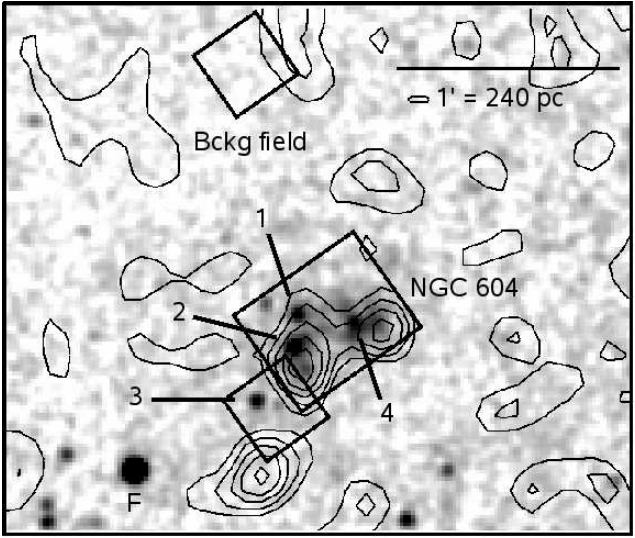}
\caption{2MASS {\em Ks}-band image of the region of NGC\,604. IR sources 
  labeled as 1, 2, 3, and 4 are associated directly with the molecular clouds
  (contours, adapted from Engargiola et al. 2003). Fields observed with the NICMOS
  camera 2 are marked as rectangles in full lines. North is up and East to the 
  left. The point source labeled as ``F'' is a galactic foreground star.} 
\label{fig:fov}
\end{figure}

\section{Observations}\label{s:observations}

We are dedicating a large observational effort to do a thorough
multi-wavelength study of \object{NGC\,604}, from the far ultraviolet to the
near infrared with additional high-resolution millimetric observations,
combining existing data with the new ones. 
The new HST data are divided into three sets: (a) slitless objective-prism FUV 
spectroscopy obtained using ACS/SBC, (b) multi-filter (6 broadband and 2 
narrowband) NUV to optical imaging obtained using ACS/HRC, and (c) 
NICMOS/NIC2 broadband imaging. These datasets were
obtained under HST proposals 10419 and 10722. The existing 
data include STIS-NUV objective-prism spectroscopy and photometry obtained
under HST/GO program 9096, as well as archival HST/WFPC2 imaging. Additional
observations were obtained using Gemini North/NIRI in 3 broadband
and 3 narrowband NIR filters and with the Combined Array for Research in
Millimeter-wave Astronomy (CARMA) in multi-configurations C, D, and E, to 
observe the
CO($1\rightarrow0$) transition with a velocity resolution of 3\,km\,s$^{-1}$,
and a beam size ranging from $3\arcsec$-$5\arcsec$.

The main goals of this project are: to obtain spectral classifications for 
about 200 stars, measuring their spectral energy distribution (SED) from 
130\,nm to 2.2\,$\mu$m, to identify embedded very young stellar populations 
hidden inside dust clouds, to measure the extinction law and its possible 
variations as a function of the environmental conditions, and to analyze the 
relationship between the hot stars and the surrounding gas. 
The combination of all data will allow to
obtain the most complete and detailed extinction corrected Hertzsprung-Russell
diagram of a resolved SOBA. The SEDs obtained will be analyzed with CHORIZOS
\citep{chorizos} as the primary code.

%
\begin{figure*}
\centerline{\includegraphics[width=10cm]{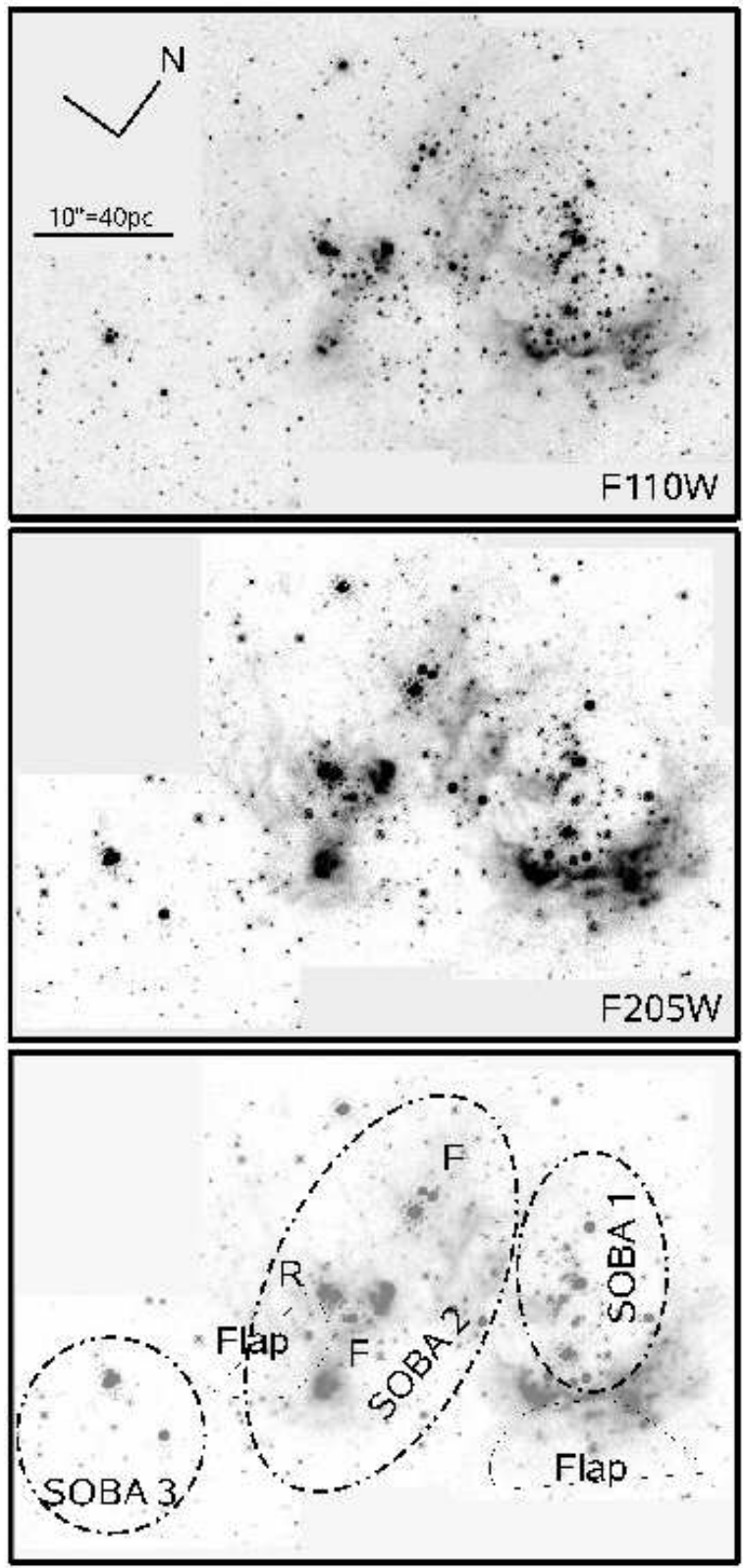}}
\caption{NICMOS F110W (top panel) and F205W (middle panel) mosaics of
  NGC\,604. The bottom 
  panel is an explanatory diagram for some structures discussed in the
  text. SOBAs labeled as ``1'', ``2'' and ``3'' are the main SOBA, the 
secondary SOBA and a third one located to the south, respectively. 
The ``F'' letter indicates the optically visible gaseous filaments. 
Flaps are areas with high optical extinction. 
``R'' refers to the red supergiant star discovered by \cite{terlevichetal96}}
\label{fig:nic2}
\end{figure*}

\section{A NICMOS view of NGC\,604}\label{s:nicmos}

Six {\em HST} orbits were requested and allocated to observe six selected
fields in and near NGC\,604 with the NICMOS camera 2 (NIC2), using the 
broadband filters F110W, F160W and F205W (similar to the {\em J}, {\em H}, and 
{\em K} passbands, respectively). Figure~\ref{fig:fov} shows the observed 
fields on a {\em Ks} image extracted from 2MASS. 
A $37\arcsec\times32\arcsec$ ($\sim140$\,pc) area centered on the core of
\object{NGC\,604} was observed in four tiles using a dithering strategy. An
adjacent field to the south of the core, centered
on a bright NIR source was also observed. In addition, sky frames were
obtained in a nearby ($1\arcmin\equiv240$\,pc away) field for background
subtraction and to take into account the stellar population of the \object{M33}
disk in the subsequent photometric analysis.

Molecular clouds associated with \object{NGC\,604} have been detected in CO 
by \citeauthor{engargiola2003} (\citeyear{engargiola2003} and references
therein). The 2MASS {\em Ks}-band image (Fig.\,\ref{fig:fov}) shows a 
concentration of point-like sources in the core of \object{NGC\,604}, 
some of them surrounded by an extended halo.  
In contrast, the ``background'' field seems clean of such bright sources.
The 2MASS IR sources are associated directly with the molecular clouds. This
morphological relationship between infrared sources and molecular clouds
resembles the \object{30\,Dor} Nebula.

\begin{figure}[t]
\includegraphics[width=\columnwidth]{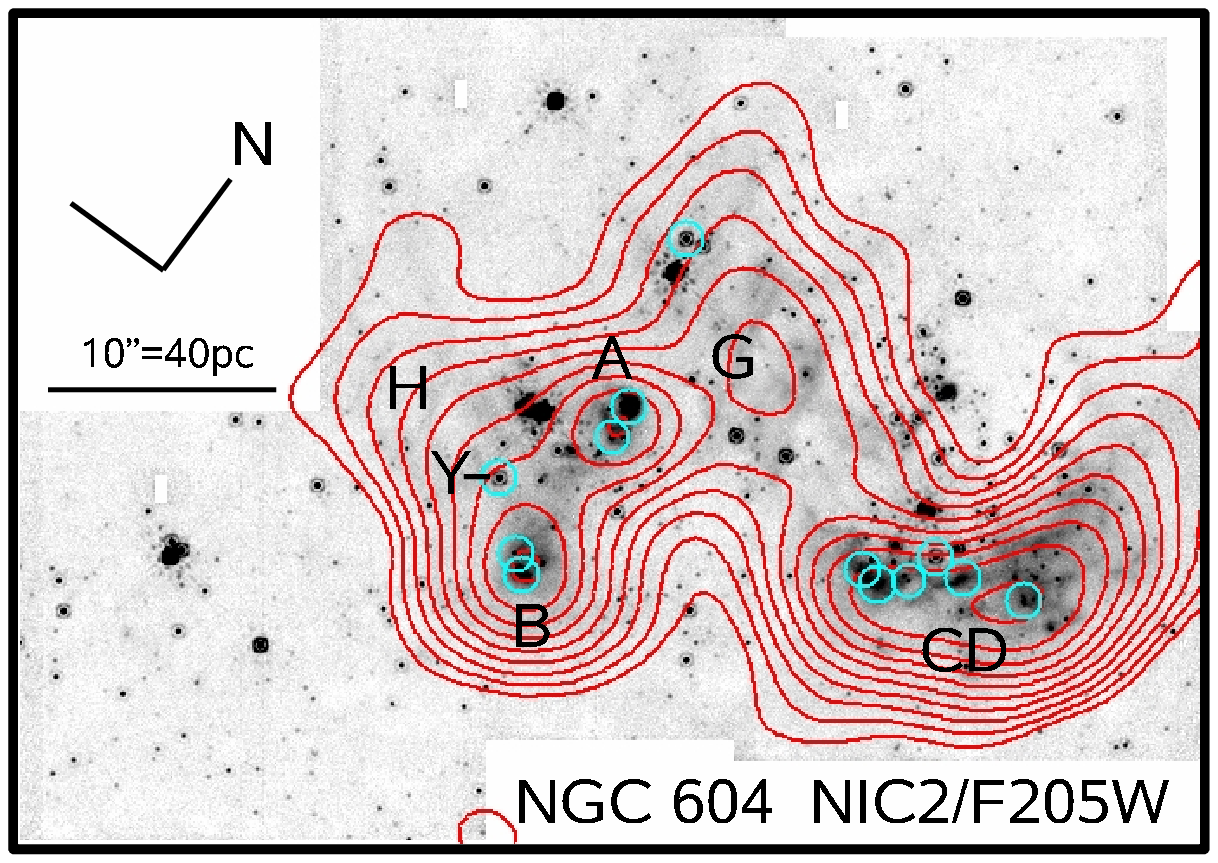}
\caption{Contour diagram of the 8.4 GHz radio continuum (adapted from
  Churchwell \& Goss 1999) superposed on the F205W NIC2 image. 
  Labels indicate the radio continuum structures (see Ma\'{\i}z Apell\'aniz et al. 2004 for
  an explanation). Circles mark the position of the massive young stellar 
object candidates}
\label{fig:radio}
\end{figure}

The new NICMOS broadband imaging of \object{NGC\,604} reveals new IR point
sources, clusters and diffuse structures, improving the spatial resolution 
of the 2MASS data by more than an order of magnitude. 
Figure~\ref{fig:nic2} shows the F110W and F205W mosaics of five NIC2 fields in
the observed area of \object{NGC\,604}. The variety of observed structures is 
striking. The two optically detected SOBAs (main and secondary) are easy to 
discern. They are encircled by labeled ellipses in
Fig.\,~\ref{fig:nic2} (see \citeauthor{maizetal2004} \citeyear{maizetal2004}
for detailed explanation of the
optical structures). An additional SOBA affected by larger extinction is
present to the south (labeled as ``SOBA 3''). 
The two cavities associated to the SOBAs are defined. Cavity A (associated to
the main SOBA) is limited by a bright rim to the south. This rim is brighter 
in the F205W filter, when compared with the F110W filter, indicating that it is
affected by a large extinction. The extended nebular emission to the
southwest in the F205W image is coincident with the structure labeled as
``flap'' by \cite{maizetal2004}, an (optical) obscuration area which is 
a dusty screen associated with the molecular cloud in that region.
This main CO cloud lies on the southern edge of the bright nebular rim. 
\cite{churchwell1999} and
\cite{maizetal2004} have analyzed the extinction in \object{NGC\,604} using
radio continuum, H$\alpha$, and H$\beta$ images. They have found a very good
correlation between the CO emission and the total optical extinction,
which indicates that the dust is associated with molecular gas. Some of the
\ion{H}{2} gas is located behind large-optical-depth well-defined dust
``flaps'' that make it invisible in the optical. Figure~\ref{fig:radio} shows
the 8.4 GHz radio continuum contours (from \citep{churchwell1999})
superimposed on the F205W image. Again, the correlation between radio and
$2.2\mu$m emission seems to be very good. 
The radio peaks C and D are superposed to the southwest NIR nebular emission. 
The radio peaks A and B are coincident with nebular knots. 
Peak A is one of the brightest H$\alpha$ structures (see 
\citep{maizetal2004}), and the same happens in the NIR. 
Gemini/NIRI images (not shown) demonstrate that most of such emission is 
produced in the warm gas as Brackett $\gamma$ emission line. 
A very interesting result is the remarkable spatial correlation between the 
massive young stellar object candidates and the radio peaks, which suggests 
that those areas are prime candidates for massive star forming regions.

%

\section{NICMOS photometry}\label{s:photo}

\begin{figure*}
\includegraphics[width=\columnwidth]{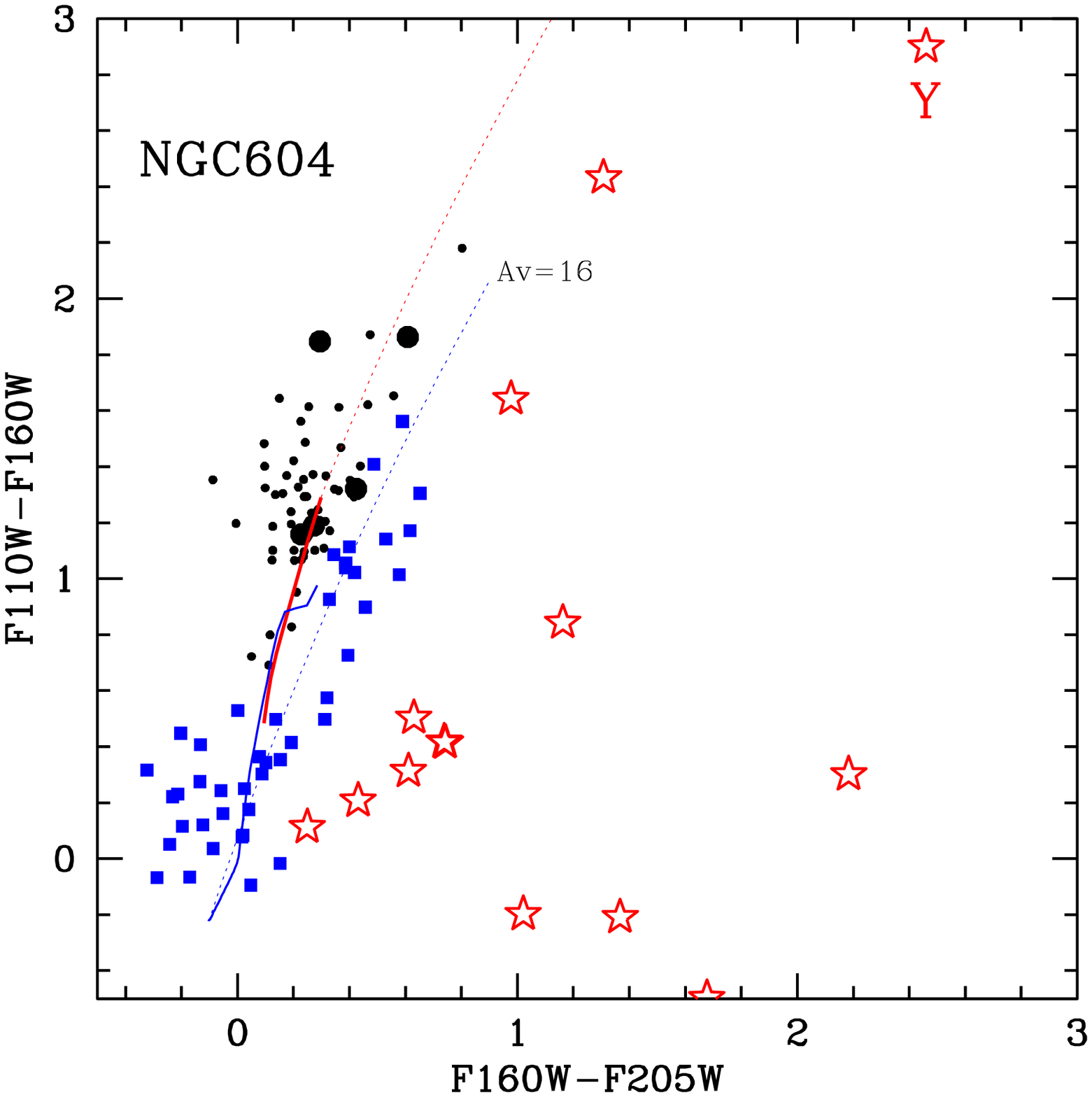}
\includegraphics[width=\columnwidth]{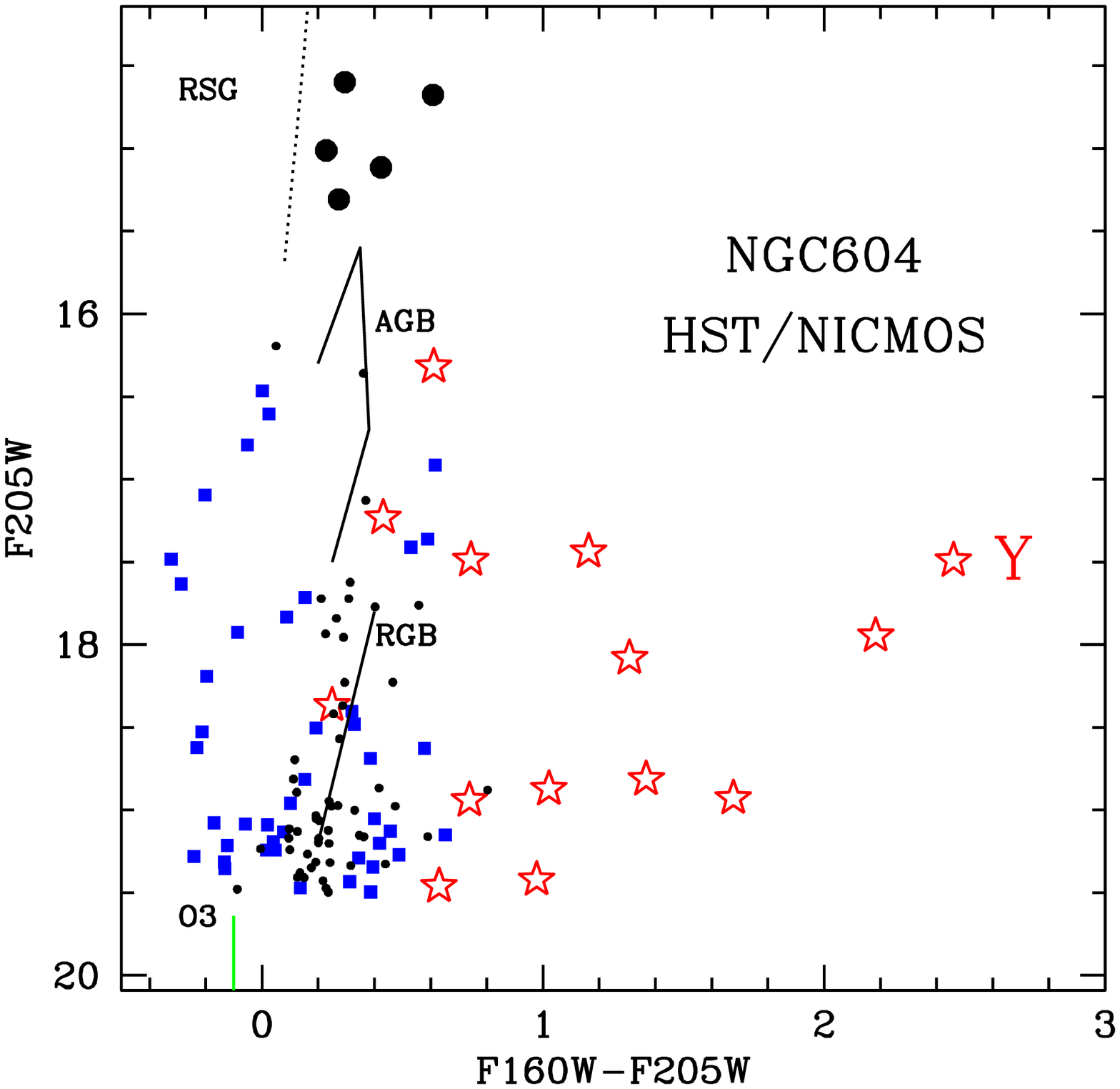}
\caption{NIC2 color-color (left) and color-magnitude (right) diagrams. This
  plot includes all sources with $m_{\rm F205W}<19.5$. Color-color diagram:
  the main-sequence locus and the cool-giant branch are indicated as solid 
  lines; reddening tracks for $A_V=16$ are also plotted. Color-magnitude
  diagram: red supergiant star locus is indicated as a dotted line in the
  upper part of the diagram; the asymptotic giant branch and red giant branch 
  are also plotted as full lines. Symbols:
  ``stars'' are MYSO candidates; ``squares'' are hot stars; ``dots'' are mostly
  red giant stars; ``black circles'' are red supergiants stars
  }
\label{fig:photo}
\end{figure*}

Point spread function photometry was performed in the NIC2 mosaics and
``background'' field.  In total, we have more than 600 detections in the three
filters with photometric errors less than 0.3 mag. 
The photometric sensitivity is about 21.9\,mag in the F205W filter, 
corresponding to an unreddened early-B type star at the distance of M33. 
Figure\,\ref{fig:photo} shows the color-color and color-magnitude diagrams 
for all sources brighter than $m_{\rm F205W}=19.5$ (O3\,V star at M33). 
We decided to preserve our data in the NICMOS photometric system in order to 
avoid additional problems with color transformations to ground-based 
photometric systems. 
Main-sequence and red giant stars loci are plotted as continuum lines in the 
color-color diagram in  Fig.\,\ref{fig:photo}. 
They are synthesized using CHORIZOS, Kurucz 
models, and NIC2 filter transmission curves. 
Reddening vectors ($A_V=16$) for an early-type star and a red giant have been 
derived in the same way, adopting a normal reddening law. 
In the color-color diagram is possible to discern three main population
of objects: (a) {\em blue}, {\em hot} stars with small o moderate reddening close to the
main-sequence locus and along the reddening track; (b) {\em red} stars in the
red giant locus; (c) a population of stars with infrared excess to the right 
of the early-type star reddening vector. 

%

The distribution of sources in the color-magnitude diagram shows that the
{\em blue} sources (group a) correspond mostly to luminous hot stars
associated to SOBAs (massive stars). Meanwhile, stars in the group (c)
are mostly massive young stellar object (MYSO) candidates and/or extreme emission
line objects. Objects in group (b) are stars belonging to the red giant 
branch, and they are more or less smoothly distributed in the mosaic. 
The brightest objects in the field are five stars in group (b).
Their positions in the color-magnitude diagram suggest that
they are red supergiant stars. In fact, one of them has been spectroscopically 
identified as such by \cite{terlevichetal96}. 
This star is a close companion to a compact massive cluster and two WR star 
candidates (WR11 and WR7, see
\citeauthor{drissenetal93} \citeyear{drissenetal93} and
\citeauthor{maizetal2004} \citeyear{maizetal2004}). 
The other four red supergiant star candidates are also located very close (few
arc-seconds) to bona-fide WR stars, WR candidates or massive hot stars. 
The spatial neighborhood observed between the RSG stars and the early-type 
objects addresses the question of whether they belong to the same star 
generation or they just represent a superposition of different populations, 
as it is the case of \object{30 Dor}. 
In the latter GHR, \cite{walbornblades97} proposed the coeval existence of 
five distinct populations, nicely distinguished by age and distribution. 
Likewise, this issue indicates the complexity of the stellar content in a 
GHR.

Among the MYSO candidates, a particular source deserves special consideration, 
owing to its large NIR colors. 
This object, labeled as ``Y'' in Fig\,{fig:radio} and
Fig.\,\ref{fig:photo} is barely detectable in the F110W image. Its
NIR photometric properties are very similar to those of the MYSO in 30 Dor
(\citeauthor{rubioetal98} \citeyear{rubioetal98}; \citeauthor{brandneretal01} 
\citeyear{brandneretal01}), indicating large luminosity. The source Y is
located in a region of very large extinction, close ($2\arcsec$) to a filled 
\ion{H}{2} region (radio peak B) and the molecular cloud core. Thus, this area
has all the necessary ingredients to be a star forming region.

\section{Summary}

In this contribution we have presented the first high-spatial resolution NIR 
images of NGC\,604, obtained with the NIC2 camera aboard {\em HST}. 
These new NICMOS broadband images reveal new NIR point sources, clusters and 
diffuse structures. We found an excellent spatial correlation between 8.4 GHz 
radio continuum and 2.2\,$\mu$m nebular emission. 
We also found that MYSO candidates appear aligned with those radio peak 
structures, reinforcing the idea that those areas are star-forming regions. 
Three different SOBAs can be recognized in the new NICMOS images. 
moreover, the brightest NIR sources in our images show properties that suggest 
the presence of RSG stars. 

This preliminary analysis of the NICMOS images shows the extreme complexity of 
the stellar content in the NGC\,604 nebula.

%
\acknowledgments
RHB would like to thank the Scientific Organizing Committee for the invitation to
participate in the meeting and the IAA staff and Director for
the warm hospitality during his visit to the Institute.
We would like to thank Greg Engargiola and Ed Churchwell
for giving us access to their data. Support for this work was provided by
FONDECYT Project No. 1050052 (RHB), by the Chilean {\sl
Center for Astrophysics} FONDAP No. 10510003 (MR), and by the Spanish Ministerio de
Educaci\'on y Ciencia (JMA and EP) through the Ram\'on y Cajal program and grants AYA2004-02703 
and AYA2007-64712, which are co-financed with FEDER funds.


%
\bibliographystyle{spr-mp-nameyear-cnd}  

%

\end{document}